%% file: main.tex
\documentclass[lettersize,journal]{IEEEtran}
\usepackage{amsmath,amsfonts}
\usepackage{algorithmic}
\usepackage{algorithm}
\usepackage{array}
\usepackage[caption=false,font=normalsize,labelfont=sf,textfont=sf]{subfig}
\usepackage{textcomp}
\usepackage{stfloats}
\usepackage{url}
\usepackage{verbatim}
\usepackage{graphicx}
\usepackage{cite}
\usepackage{booktabs}
\usepackage{colortbl}
\usepackage{xcolor}
\usepackage{multirow}
\usepackage{enumitem}
\usepackage{makecell}

\newcommand{\br}{\color{black}}

\input{notations}

\begin{document}


\title{D2ToF: One-Step Diffusion for Depth Super-Resolution from Lightweight Single-Photon LiDAR Waveforms}

\author{\IEEEauthorblockN{
Pallabjyoti Deka, Lewis Wilson, Stephen McLaughlin, Jonathan Leach and Abderrahim Halimi}\\
\IEEEauthorblockA{School of Engineering and Physical Sciences,
Heriot-Watt University, Edinburgh UK
}
\thanks{This work was supported by the Engineering and Physical Sciences Research Council (EP/Z533166/1 and EP/Z533178/1). For the purpose of open access, the author(s) has applied a Creative Commons Attribution (CC BY) license to any Accepted Manuscript version arising.}%
}




\maketitle

\begin{abstract}
Single-photon LiDAR (SPL) sensors offer exceptional temporal resolution and single-photon sensitivity for depth imaging. However, cost-effective SPL arrays typically suffer from extremely low spatial resolutions (e.g. $8\times 8$ pixels), limiting their real-world utility. In this paper, we propose a novel, one-step super-resolution framework, \textbf{D2ToF}, \textbf{D}istilled \textbf{D}iffusion for \textbf{T}ime-\textbf{o}f-\textbf{F}light, that bridges the gap between high-fidelity generative modeling and real-time execution. Our approach uses an amortized variational distillation strategy that allows for a measurement-based data consistency term to be incorporated directly into the one-step training objective. This encourages that the synthesized high-resolution depth maps are consistent with the captured low-resolution Time-of-Flight (ToF) histograms.
To overcome the severe ill-posedness of upsampling extremely degraded inputs, we introduce a multimodal guidance strategy. By leveraging a foundational monocular depth model to extract structural priors from a high-resolution reflectivity image, our network is guided by a rich triad of the ToF histogram, reflectivity map, and relative depth estimate. Experimental results outperform evaluated baselines, achieving $16\times$ spatial super-resolution (from $8\times 8 \times 100$ histograms to $128\times 128$ depth maps) at an accelerated inference speed of approximately $83$ milliseconds.
\end{abstract}



\input{sections/01_introduction}
\input{sections/02_background}

\input{sections/03_methodology_v3}

\input{sections/04_experiments_v2}
\input{sections/05_conclusion}


\bibliographystyle{IEEEtran}
\bibliography{ref}

\vfill


\end{document}

%% file: notations.tex
\input com_notations.tex

\newcounter{algo}
\renewcommand{\thealgo}{\arabic{algo}}

%% file: com_notations.tex
\def\bsc{{\boldsymbol{c}}}
\def\bsd{{\boldsymbol{d}}}

\def\bsh{{\boldsymbol{h}}}

\def\bss{{\boldsymbol{s}}}

\def\bsw{{\boldsymbol{w}}}
\def\bsx{{\boldsymbol{x}}}
\def\bsy{{\boldsymbol{y}}}

\def\bsA{{\boldsymbol{A}}}

\def\bsI{{\boldsymbol{I}}}



%% file: sections/01_introduction.tex
\section{Introduction}\label{sec:intro}

Depth imaging supports a broad range of applications including 3D reconstruction for autonomous navigation~\cite{Rapp_JSPM_2020,Wallace_TVT2020} and consumer electronics~\cite{low-cost-clara}. Among
depth-sensing technologies, single-photon LiDAR (SPL) is attractive for its exceptional temporal resolution and single-photon sensitivity~\cite{Pawlikowska_OE2017,Li_Optica_21}, which enable accurate ranging even at long range and under low illumination. Cost-effective lightweight SPL sensors however, provide only a small number of pixel waveforms; typically $8\times8$ or $16\times16$~\cite{Ruget_OE25_PnP, Ruget_SA2023}. This severe spatial limitation restricts their practical use and motivates computational super-resolution (SR) approaches to recover a high-resolution depth map from such measurements.

\begin{figure}[!htb]
    \centering
    \includegraphics[width=0.9\linewidth]{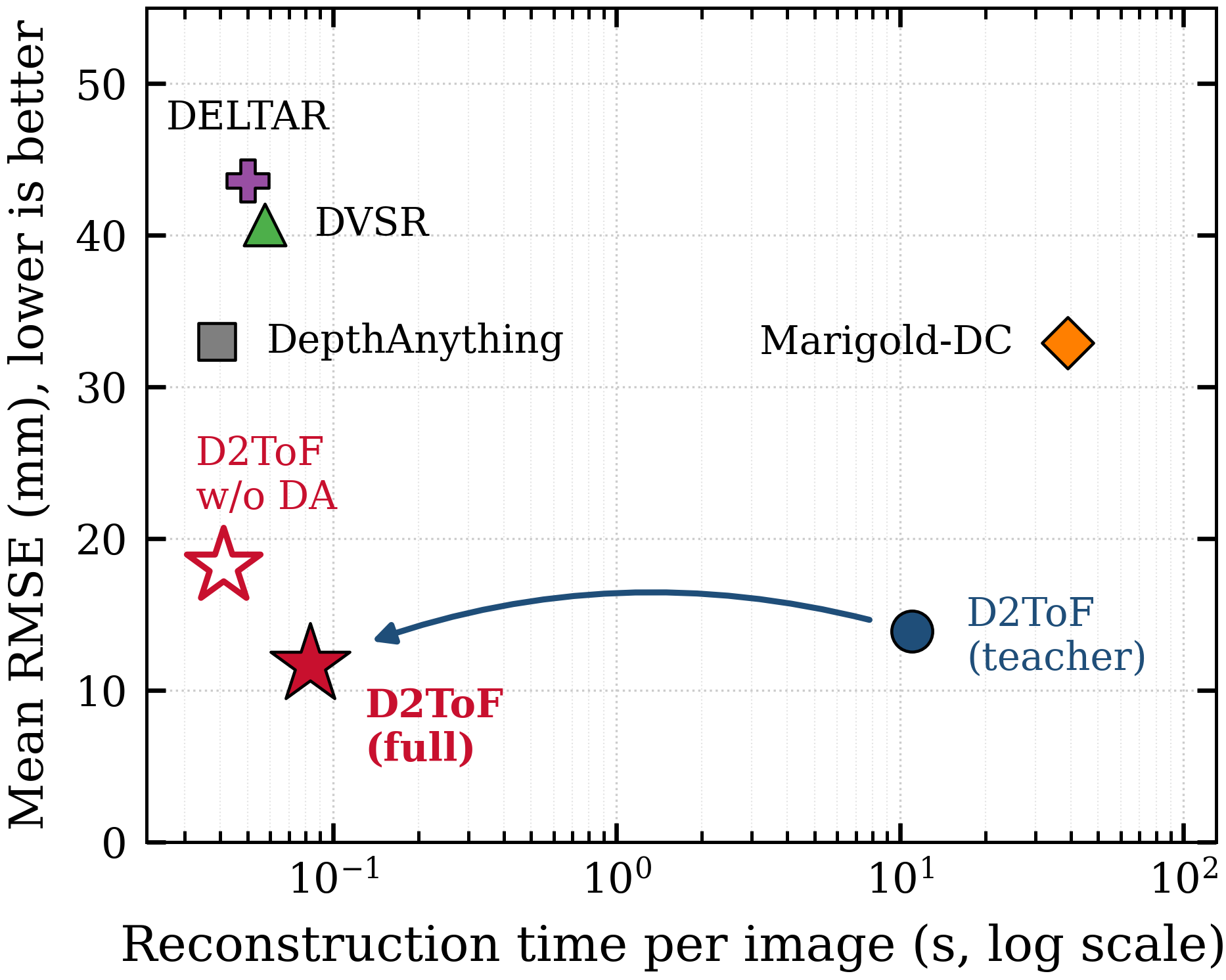}
    \caption{Reconstruction quality (RMSE in mm) vs. time (sec.) of our proposed method in comparison to DELTAR~\cite{deltar}, DepthAnything-V2~\cite{depthanythingv2}, DVSR~\cite{sun2023consistent}, and Marigold-DC~\cite{marigold-dc} evaluated on 12 clean scenes of Middlebury dataset.}
    \label{fig:speed-quality}
\end{figure}

{\br
To compensate for the low spatial resolution, sensor-fusion methods use a co-registered high-resolution (HR) intensity or reflectivity image to guide the upsampling of low-resolution (LR) LiDAR data~\cite{Lindell_sp,Ruget_OE21,sun2023consistent,gyongy2020high,Sun_OE2020,Kang_OE2022,callenberg2021super,Halimi_SR2026,deka_ssp}. These methods span statistical reconstruction~\cite{halimi2021robust,Rapp_TCI2017,choi2025graph}, learning-based feature fusion~\cite{Lindell_sp,Ruget_OE21,Ruget_OE25_PnP,sun2023consistent}, and, more recently, to generative models using conditional diffusion~\cite{deka_ssp}. Although these approaches improve reconstruction quality, plug-and-play methods~\cite{Ruget_OE25_PnP} and iterative diffusion~\cite{deka_ssp, marigold-dc} incur substantial inference costs (Figure~\ref{fig:speed-quality}). Recovering detailed depth from a single frame becomes particularly challenging when the LiDAR provides only $8\times8$ waveforms (Figure~\ref{fig:real-plot}). Earlier approaches exploit multiple frames from a video sequence~\cite{Ruget_OE25_PnP,sun2023consistent} or operate at a higher base spatial resolution~\cite{deka_ssp}, leaving efficient reconstruction from a single frame at such low spatial resolution insufficiently explored.
}

We address both obstacles with a framework built on amortized variational distillation. We first learn a multi-step conditional diffusion teacher over HR depth maps and then distil this distribution into a one-step generator. Following the amortized formulation of~\cite{davi}, a data-fidelity term is integrated to enforce consistency of the reconstruction with the observations. We re-render the predicted HR depth back into a LR ToF histogram and penalise its deviation from the measured histogram. This term anchors the reconstruction to the acquired data and, as our ablations show, improves robustness when the guiding image is degraded.

Reconstructing from an $8\times8$ waveform also requires strong spatial priors, since the histogram itself carries limited layout. An HR reflectivity image supplies fine texture but is inherently ambiguous about 3D geometry. We therefore add a monocular depth guidance: we pass the reflectivity image through a frozen foundational depth model
(e.g. DepthAnything\,V2~\cite{depthanythingv2}) to obtain a relative depth estimate, and condition the network on the resulting triad of ToF histogram, reflectivity map, and monocular depth. The monocular depth estimator adds a small, fixed cost at inference while contributing dense geometric structure, object boundaries and surface layout that the raw waveform lacks.

We evaluate the method on simulated Middlebury scenes at different photon levels and on real captures from a cost-effective $8\times8$ STMicroelectronics SPAD sensor, including low-light acquisitions. Across these settings the proposed one-step model outperforms the learning-based~\cite{deltar,sun2023consistent,marigold-dc}, and monocular baselines~\cite{depthanythingv2} we compare against, while reconstructing a $128\times128$ depth map in approximately $83$\,ms about $130\times$ faster than the multi-step teacher it is distilled from, while achieving lower average RMSE on the reported benchmark.

In summary, our main contributions are as follows:

\begin{itemize}
  \item We propose a one-step framework for extreme single-photon depth super-resolution that reconstructs a $128\times128$ depth map from an  $8\times8$ histogram in a single forward pass (${\approx}83$\,ms),  roughly $130\times$ faster than the {\br 250-step (DDPM)} teacher it is distilled from.
  
  \item We introduce a differentiable histogram data-consistency term, derived from the single-photon observation model and embedded in the amortized objective. This anchors the reconstruction towards the measured  histogram.
  
  \item We propose a multimodal conditioning strategy that combines the  ToF histogram, an HR reflectivity image, and a frozen foundational monocular-depth estimate, enabling the recovery of geometric structure  that is absent from the raw $8\times8$ waveform.
  
  \item We validate the method on simulated and real $8\times8$ SPAD data, including low-light conditions, showing strong performance and robustness to degraded reflectivity guidance without retraining.
\end{itemize}

The remainder of this paper is organised as follows. Section~\ref{sec:background} reviews the single-photon observation model and the background on diffusion models and distillation that our method builds upon. Section~\ref{sec:method} presents the proposed framework: the conditional diffusion teacher, the multimodal conditioning strategy, the network architecture, and the one-step distillation framework with its histogram data-consistency term. Section~\ref{sec:experiments} reports results on simulated Middlebury scenes and on real captures from an $8\!\times\!8$ SPAD sensor, and Section~\ref{sec:exp-ablation} presents the ablation studies. Section~\ref{sec:conclusion} concludes the paper. Code will be available upon acceptance.

%% file: sections/02_background.tex
\section{Background}
\label{sec:background}

\subsection{Single-photon LiDAR Observation model} \label{subsec:Observation_model}

\begin{figure}
    \centering
    \includegraphics[width=1\linewidth]{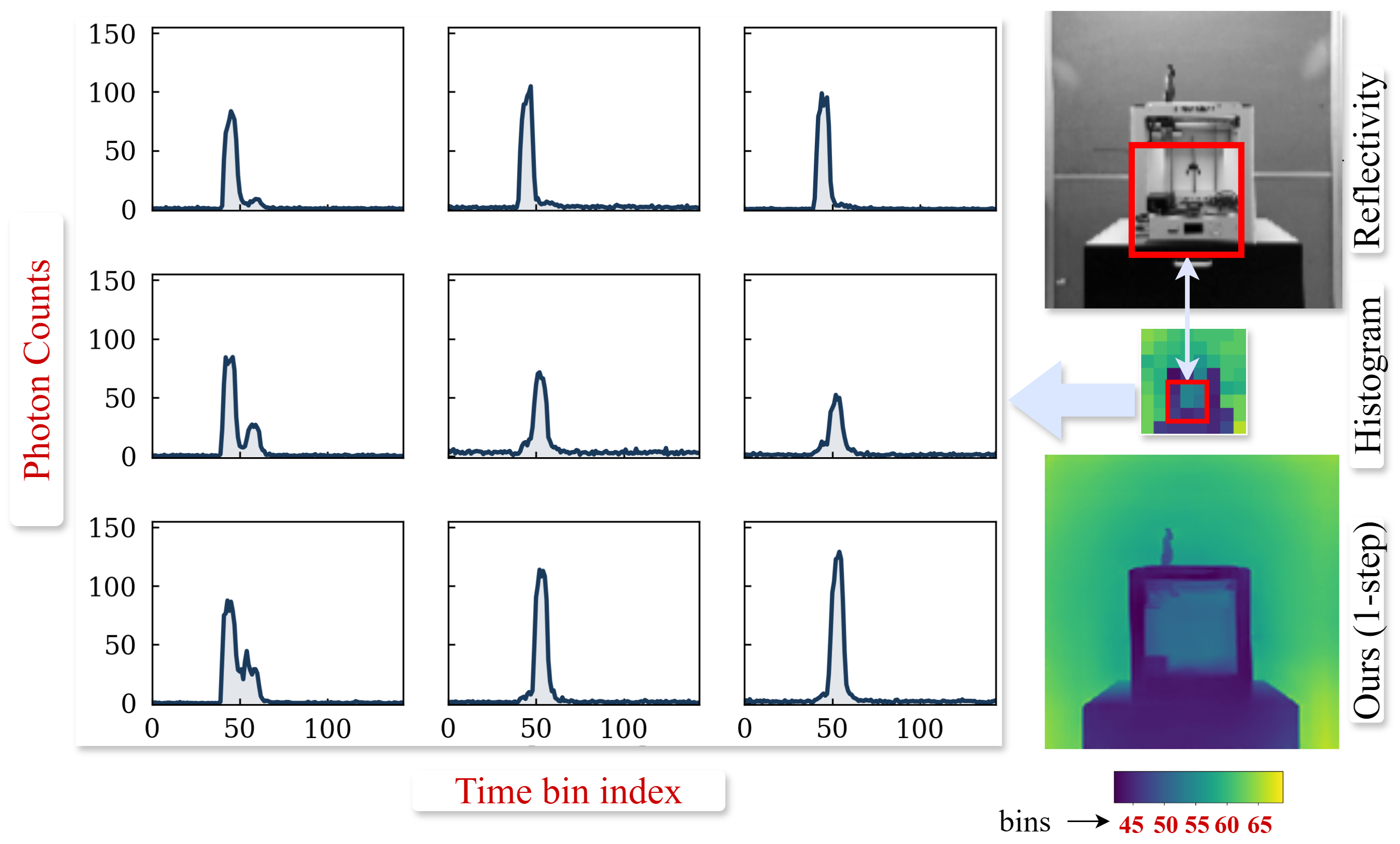}
    \caption{Real \(8{\times}8\) SPAD measurement and corresponding reconstruction. Right: (top) high-resolution guidance image with the selected field of view, (middle) spatial locations of the selected \(3{\times}3\) ToF cells within the measured \(8{\times}8\) array, (bottom) one-step reconstructed depth. Left: photon-arrival waveforms for the  highlighted ToF cells, showing the temporal data available.}
    \label{fig:real-plot}
\end{figure}

A LiDAR system emits light pulses and records the ToF of reflected photons across pixel locations. For each pixel, a  histogram waveform of photon arrivals is constructed with \(h_{n,k}\) denoting the photon count at pixel \(n \in \{1,\ldots,N\}\) and time bin \(k \in \{1,\ldots,K\}\).
The histogram waveform $\bsh_{n}=\{h_{n,k}, \forall k\} $ follows a Poisson distribution \cite{Halimi_TCI2020,Rapp_TCI2017}:

\begin{equation}
    h_{n,k} \sim \mathcal{P}(s_{n,k} ), \textrm{~ with ~~ } s_{n,k} = c_{n} f(k-d_n) + b_{n} 
    \label{eqt:Statistical_model_init}
\end{equation}

where $\bsd=\{d_{n}, \forall n\}, \bsc=\{c_{n}, \forall n\}$ are the target's depth and reflectivity parameters, $f$ represents the  system impulse response function (IRF) assumed Gaussian with variance $\sigma^2$, and $b_{n}$ denotes the background counts due to the observation environment or detector's dark counts.

Given the histogram waveforms $\bsh_{n}, \forall n$, the goal is to estimate the target's depth maps $\bsd$ to reconstruct the 3D scene. A classical strategy is the log-matched filter algorithm,  equivalent to the maximum likelihood estimator (MLE) in absence of background (i.e., $b_n=0$), given by

\begin{equation}
\hat{d}_n = \max_d \sum_{k=1}^{K}  h_{n,k}   \log \left[f(k-d)\right].
\label{eqt:Matched_Filter}
\end{equation}

\subsection{Observation model for cost-effective SPAD sensors} \label{subsec:cost-eff-obs-model}
Recent advances in single-photon technology have enabled the development of SPAD arrays for fast imaging. However, cost-effective sensors often only provide a limited number of pixel waveforms (e.g., $8 \times 8$), resulting in low-resolution images where each pixel covers a large surface of the scene. This impacts the observation model and can be expressed as
\begin{equation}
    \bsh_{:,k} \sim \mathcal{P}\!\left( \bsA \bss_{:,k} \right),  \forall k
    \label{eqt:Statistical_model}
\end{equation}
where $\bsA$ denotes a spatial convolution–decimation operator modeling the limited spatial resolution of the sensor. In such cases, the classical log-matched filter algorithm performs poorly due to the presence of multiple signal peaks in each waveform, motivating the need for advanced algorithms that exploit full LiDAR waveform for improved depth estimation.


In the dense-photon regime, Poisson count statistics can be approximated by Gaussian statistics~\cite{anscombe,reweighted_l2}. Following the Gaussian histogram models adopted for SPL in~\cite{eusipco_2023,Ruget_OE25_PnP}:

\begin{equation}
    \bsh_{:,k} = \bsA\bss_{:,k} + \mathbf{e}_k, \qquad
    \mathbf{e}_k \sim \mathcal{N}(\mathbf{0}, \eta\mathbf{I}),
    \label{eqt:Gaussian_model}
\end{equation}

where \(\eta\) denotes a noise variance. Under this model, the negative log-likelihood reduces to a squared \(\ell_2\) distance, which we adopt as the data-fidelity term in Sec.~\ref{sec:method}.


The high-resolution reflectivity image $\mathbf{R} \in [0,1]$ that guides the reconstruction is acquired by a conventional camera and is therefore itself photon-limited, particularly in low-light conditions. We model its degradation with the standard Poisson-Gaussian model~\cite{foi_poisson_gaussian_modeling, read_noise_brooks2019unprocessing, Chan_Li2021PhotonLimitedOD},

\begin{equation}
    {\mathbf{R}}_\text{noisy} = \frac{1}{\gamma} \mathcal{P}\!\left( \gamma \mathbf{R} \right)
    + \boldsymbol{e}_r, \qquad
    \boldsymbol{e}_r \sim \mathcal{N}\!\left( \mathbf{0}, \sigma_r^2 \mathbf{I} \right),
    \label{eqt:Reflectivity_noise}
\end{equation}

where $\gamma$ denotes the effective photon-count level and controls the signal-dependent shot noise (smaller $\gamma$ yields stronger noise), while $\sigma_r$ denotes the standard deviation of the signal-independent read noise $\mathbf{e}_r$. In Sec.~\ref{sec:experiments}, the networks are trained using clean guidance images only, while ${\mathbf{R}}_\text{noisy}$ is introduced exclusively at test time to evaluate robustness to degradation of the guidance modality.



\subsection{Diffusion Models}   \label{subsec:bg-diffusion}

Diffusion models~\cite{ddpm, score-sde} are formulated around a forward noising process, denoted $\{\bsx_t\}_{t=0}^T$, which is governed by a stochastic differential equation (SDE). This process operates over a continuous time interval $t \in [0, T]$, starting with a sample from the true data distribution $x_0 \sim p(\bsx_0)$ at $t=0$. As time progresses, the SDE gradually perturbs the data, ultimately transforming it into a sample from a simple prior, $\bsx_T \sim \mathcal{N}(0, \bsI)$, at $t=T$, where $\bsI$ denotes identity matrix.
The general form of this forward SDE is:
\begin{equation}
d \bsx = f_t \bsx_t dt + g_t d\bsw, \label{eq:fwd_sde}
\end{equation}
where $f_t : \mathbb{R} \to \mathbb{R}$ is the drift coefficient, $g_t \in \mathbb{R}$ is the diffusion coefficient, and $\bsw$ represents a standard Wiener process. 
A corresponding reverse-time SDE exists, which generates data by starting from the prior $p(\bsx_T)$ and evolving backward to $t=0$. This generative SDE is defined by:
\begin{equation}
d\bsx = [f_t \bsx_t - g_t^2 \nabla_{\bsx_t} \log p(\bsx_t)] dt + g_t d\bar{\bsw}, \label{eq:rev_sde}
\end{equation}
where $d\bar{\bsw}$ is a standard Wiener process running in reverse time and $dt$ signifies an infinitesimal negative time step. This reverse process is critically dependent on the score function, $\nabla_{\bsx_t} \log p(\bsx_t)$, which is unknown. Therefore, it is approximated using a time-conditioned neural network, $s_\theta(\bsx_t, t)$, such that $\nabla_{\bsx_t} \log p(\bsx_t) \approx s_\theta(\bsx_t, t)$. The network is trained by minimizing a denoising score matching objective:
\[
\mathbb{E}_{p(\bsx_t|\bsx_0), t \sim U[0, T]} [\lambda(t) \|s_\theta(\bsx_t, t) - \nabla_{\bsx_t} \log p(\bsx_t|\bsx_0)\|_2^2].
\]

In this objective, $\lambda(t)$ is a positive weighting function, and $p(\bsx_t|\bsx_0)$ is the Gaussian transition kernel describing the perturbation process from $\bsx_0$ to $\bsx_t$.


\subsection{Distribution Matching Distillation of Diffusion Models}   \label{subsec:bg-distillation}

Distillation~\cite{dmd-one-step-yin} has been widely explored not only for model compression but also for domain adaptation. For instance, Score Distillation Sampling (SDS) in DreamFusion \cite{dreamfusion} adapts pretrained 2D diffusion models to the 3D generation domain through a distillation process. More recently, Variational Score Distillation (VSD)\cite{prolificdreamer} has demonstrated that a variational framework can provide a more principled and effective formulation for such adaptation tasks, enabling robust knowledge transfer between domains.
Distillation has also been employed for accelerating diffusion inference~\cite{progressive, dmd-one-step-yin}. Diff-Instruct \cite{diffinstruct} distilled the multi-step diffusion process into a single-step network, dramatically improving inference speed. It introduced  Integral KL (IKL) divergence as an effective and numerically stable metric for aligning diffusion model scores, outperforming the conventional KL divergence in cases where the model and target distributions have misaligned supports.
By combining the variational framework with the IKL metric, DAVI \cite{davi} extended the concept of distillation to posterior estimation in inverse problems. DAVI learns a parameterized distribution $q_\phi(x)$ that approximates the Bayesian posterior $p(\bsx|\bsy)$ integrating  a deep diffusion prior. This formulation jointly accounts for data fidelity and learned priors using the IKL divergence, leading to a stable and efficient inference framework that amortizes the cost of posterior sampling $p(\bsx|\bsy)$ via parameterized variational distribution.

%% file: sections/03_methodology_v3.tex
\section{Methodology}\label{sec:method}

\begin{figure*}[!htb]
    \centering
    \includegraphics[width=1.0\linewidth]{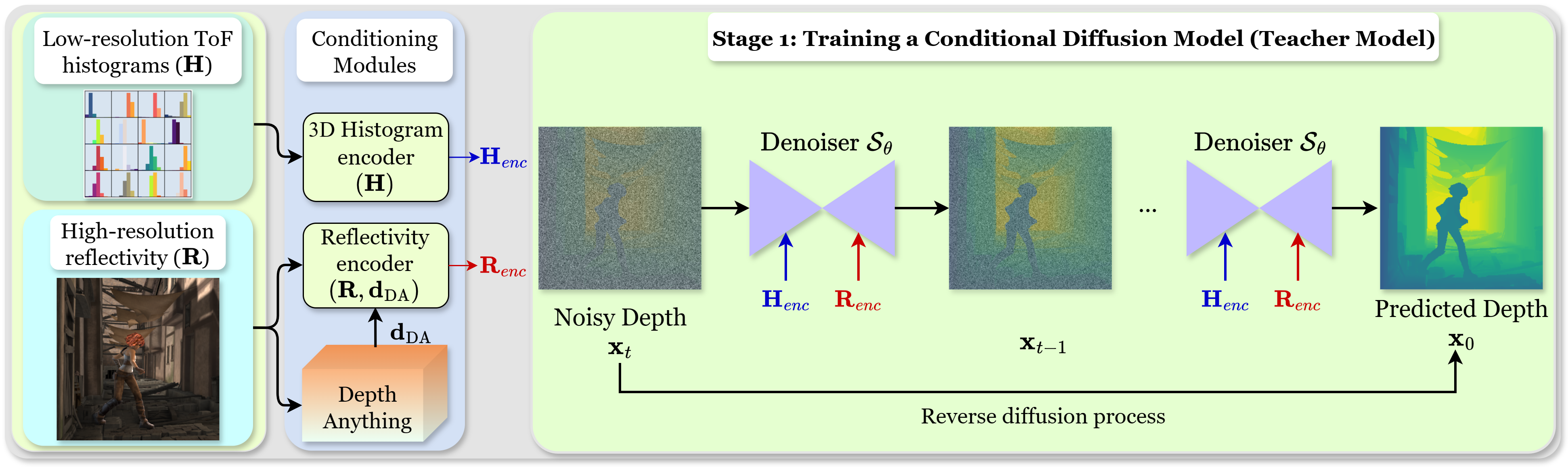}
    \caption{Overview of the conditional prior (Stage~1). The three conditioning signals: low-resolution ToF histogram $\mathbf{H}$,
    high-resolution reflectivity $\mathbf{R}$, and monocular depth prior $\mathbf{d}_{\mathrm{DA}}$ enter the U-Net asymmetrically:
    the encoded histogram $\mathbf{H}_\text{enc}$ is injected on the encoder side, while the encoded reflectivity and monocular-depth features $\mathbf{R}_\text{enc}$ are fused into every decoder (synthesis) stage.}
    \label{fig:prior_diagram}
\end{figure*}
\begin{figure*}[!htb]
    \centering
    \includegraphics[width=0.9\linewidth]{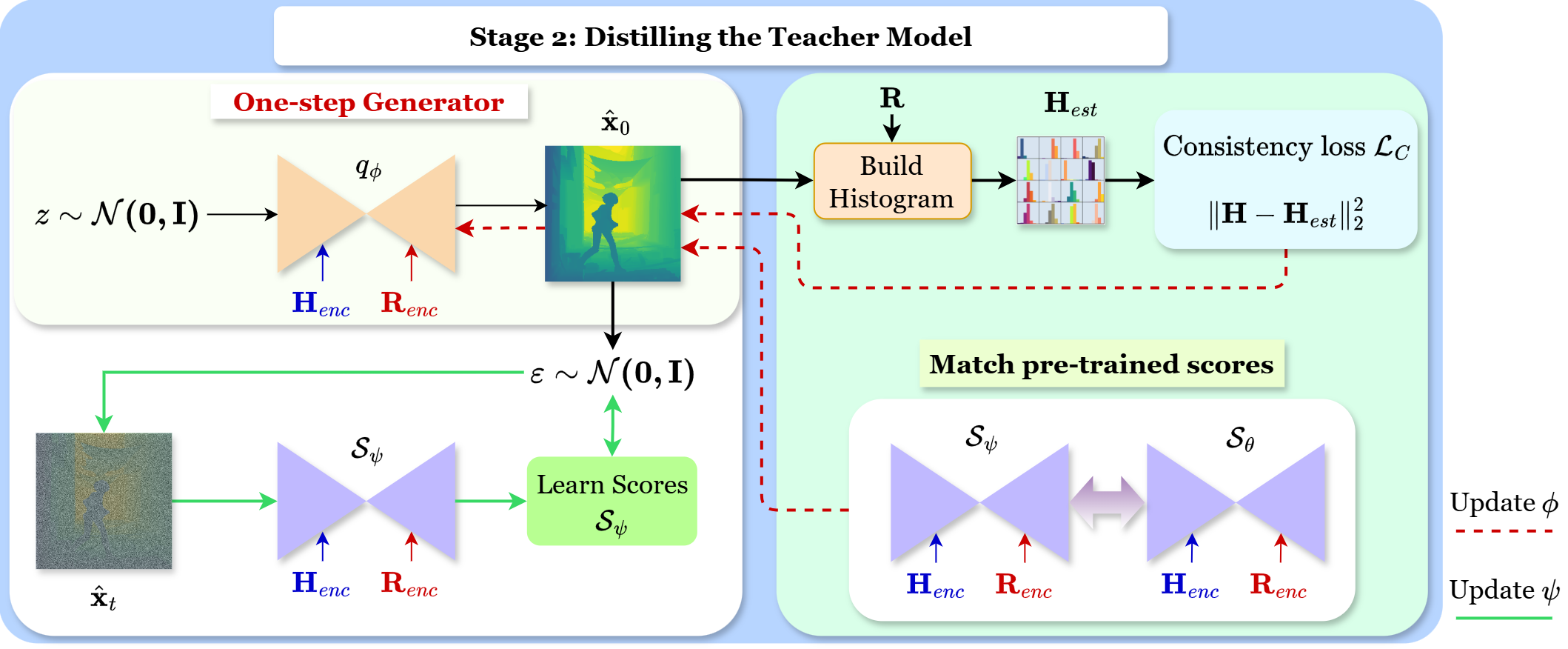}
    \caption{Overview of the distillation stage. A one-step generator $q_\phi(\mathbf{x}_0\mid\mathbf{y})$ is trained so that its samples (i) match the learned prior via teacher-student score matching and (ii) enforce estimated histogram $\mathbf{H}_{\text{est}}$ to be consistent with the measured histogram $\mathbf{H}$.}
    \label{fig:distill_diagram}
\end{figure*}

We aim to recover a $128\!\times\!128$ depth map from an $8\!\times\!8\!\times\!K$ SPAD measurement and a $128\!\times\!128$ reflectivity guidance, which is severely ill-posed: the sensor provides only $64$ spatial waveforms, and we aim to recover the missing spatial structure. Our key observation is that a faithful depth map reconstruction must satisfy two requirements simultaneously: it must (i) be consistent with the corresponding histogram when passed through the same forward model, and (ii) supply a high-frequency spatial structure that is consistent with the high-resolution reflectivity guidance. 
In addition, we focus on improving computational efficiency to support practical deployment of low-cost SPAD-based sensing platforms (such as smartphones, or robotic sensors) in the future. 
We meet these requirements as follows.
We address the extreme single-photon depth super-resolution problem with a two-stage framework.
In Stage~1 we learn a histogram-conditioned diffusion prior over high-resolution depth maps (\ref{sec:teacher}) (see Figure~\ref{fig:prior_diagram}). In Stage~2 we distil this multi-step prior into a single-step amortized posterior sampler (\ref{sec:student}) (see Figure~\ref{fig:distill_diagram}); crucially, the amortized formulation allows us to incorporate a differentiable data-consistency operator derived from the single-photon forward model, which is what grounds the generator in the measured waveforms. The result is a single forward pass that reconstructs a $128\!\times\!128$ depth map in approximately $83$\,ms, while providing much better accuracy than recent monocular depth estimation methods.

\subsection{Problem Formulation}\label{sec:method-problem}

Let $\mathbf{x}\!\in\!\mathbb{R}^{H\times W}$ denote the target depth
map at high resolution ($H\!{=}\!W\!{=}\!128$), and $\mathbf{R}\!\in\!\mathbb{R}^{H{\times} W}$ a co-registered HR reflectivity image obtained from a passive camera
aligned with the SPAD array. A cost-effective SPAD measures $M\!{=}\!M_h\!\times\!M_w$ waveforms ($M_h\!{=}\!M_w\!{=}\!8$), each recording photon counts across $K$ time bins, giving a histogram tensor $\mathbf{H}\!\in\!\mathbb{R}_{\ge 0}^{M_h\times M_w\times K}$.

Rather than relying only on the low-resolution histogram, the reconstruction is conditioned on all available observations.

%
\begin{equation}\label{eq:posterior}
  p(\mathbf{x}\mid\mathbf{y}),\qquad  \mathbf{y}=\{\mathbf{H},\,\mathbf{R},\,\mathbf{d}_{\mathrm{DA}}\},
\end{equation}
where $\mathbf{d}_{\mathrm{DA}}$ is a monocular-depth estimator output (see~\ref{sec:method-conditioning}).

\subsection{Multimodal Conditioning for Extreme Super-Resolution}\label{sec:method-conditioning}

To complement the limited information available in the low-resolution $8\!\times\!8$ photon histograms, the reconstruction of a $128\!\times\!128$ depth map is conditioned on three complementary signals, each providing a different type of information:

\begin{enumerate}
  \item \textbf{ToF histogram} $\mathbf{H}\!\in\!\mathbb{R}^{8\times 8\times K}$:
    absolute, metric depth via photon-arrival statistics, but limited spatial layout.
  \item \textbf{Reflectivity map} $\mathbf{R}\!\in\!\mathbb{R}^{H\times W}$:
    dense high-resolution spatial structure (edges, textures), but no depth.
  \item \textbf{Monocular depth estimate} $\mathbf{d}_{\mathrm{DA}}\!\in\!\mathbb{R}^{H\times W}$:
    a \emph{relative} depth field obtained by using $\mathbf{R}$ (replicated to three channels) as an input to a frozen DepthAnything\,V2~\cite{depthanythingv2} (ViT-B). This map provides dense geometric and semantic information at full resolution, including object boundaries and surface orientations.
\end{enumerate}

The three signals are complementary. The histogram provides depth but limited spatial layout, and the reflectivity provides spatial detail but no depth. Since a 2D reflectivity image is inherently ambiguous about 3D geometry, we add the monocular depth estimate to inject the geometric structure that the reflectivity alone cannot resolve. DepthAnything provides only relative disparity (scale- and shift-ambiguous). We therefore use $\mathbf d_{\mathrm{DA}}$ as unaligned geometric guidance, allowing the network to combine its spatial structure with the absolute depth information provided by the ToF histogram. We quantify the contribution of each conditioning signal and its robustness to degraded reflectivity in Sec.~\ref{sec:exp-ablation}.

\subsection{Stage~1: {\color{black}Conditional Diffusion Prior}}\label{sec:teacher}

We first learn a conditional prior to generate a high-resolution depth maps while using the observations $\mathbf{y}=\{\mathbf{H},\mathbf{R},\mathbf{d}_{\mathrm{DA}}\}$.
We consider a classical diffusion-based framework where the forward process corrupts a clean depth map $\mathbf{x}_0$ with Gaussian noise over $T\!=\!1000$ steps under a cosine schedule~\cite{nichol2021improved_t_emb}, $\mathbf{x}_t=\alpha_t\mathbf{x}_0+\sigma_t\boldsymbol{\epsilon}$, and the theta-parametrized network $\boldsymbol{\epsilon}_\theta(\mathbf{x}_t,t,\mathbf{y})$ conditioned on $\mathbf{y}$ is trained using denoising score matching:

\begin{equation}\label{eq:teacher_loss}
  \mathcal{L}_{\mathrm{teacher}}
  = \mathbb{E}_{\mathbf{x}_0,\,t,\,\boldsymbol{\epsilon}}
    \bigl\lVert
        \boldsymbol{\epsilon}_\theta\!\left(
        \alpha_t\mathbf{x}_0+\sigma_t\boldsymbol{\epsilon},\;t,\;\mathbf{y}
      \right)- \boldsymbol{\epsilon}
    \bigr\rVert_2^2 .
\end{equation}

The resulting model is sampled using $250$ DDPM steps~\cite{ddpm}. Although it produces high-fidelity depth reconstructions, iterative sampling introduces substantial inference latency. In our framework, this model serves as the multi-step teacher whose conditional score given by $\mathcal{S}_{\theta}(\mathbf{x}_t,t,\mathbf{y}) \approx \nabla_{\mathbf{x}_t} \log p_{\theta}(\mathbf{x}_t\mid\mathbf{y})$ is used for distribution matching during Stage~2. The following sections introduce the considered network architecture and the distillation process of this teacher  into a one-step generator while incorporating explicit histogram data consistency.

\subsection{Network Architecture}\label{sec:architecture}

Both teacher and student use a conditional U-Net~\cite{UNet} built on the ADM backbone~\cite{dhariwal2021diffusion} operating on $128\!\times\!128$ inputs, with two residual blocks with group normalization per resolution level carrying the sinusoidal timestep embedding. Starting from $128$ base channels, the channel width is scaled by $(1,1,2,3,4)$ across the five resolution levels ($128\!\times\!128$ to $8\!\times\!8$), giving $128$ to $512$ channels as the spatial resolution decreases. Multi-head self-attention is applied at the two coarsest levels, $16\!\times\!16$ and $8\!\times\!8$.

To effectively leverage the complementary nature of our conditioning signals, we employ an asymmetric injection strategy. Following~\cite{Lindell_sp}, the low-resolution ToF histogram ($\mathbf{H}$) is injected into the encoder to guide the coarse spatial scales, while the high-resolution reflectivity ($\mathbf{R}$) and DepthAnything output ($\mathbf{d}_{\mathrm{DA}}$) are fed into the decoder stages to synthesize fine structural details. Our UNet consists of $140$M parameters including both conditioning encoders.

\subsubsection{Histogram injection (encoder side)}
Before injection, $\mathbf{H}$ is processed by a 3D convolutional encoder consisting of two residual blocks with $3\times3\times3$ kernels and channel dimensions $1\!\to\!32\!\to\!64$. To raise the effective photon count per bin in the photon-sparse regime, we first sum every four consecutive time bins, reducing $K\!=\!100$ to $25$ aggregated bins - trading marginal temporal resolution for higher signal-to-noise while preserving depth discrimination. The temporal dimension is then flattened into channel features and mapped via 2D residual blocks. These resulting features are added into the network at its two coarsest stages: the native $8\!\times\!8$ resolution and a bicubically upsampled $16\!\times\!16$ resolution.

\subsubsection{Reflectivity and prior fusion (decoder side)}
Conversely, the HR reflectivity $\mathbf{R}$ and its corresponding DA output $\mathbf{d}_{\mathrm{DA}}$ are passed through parallel hierarchical 2D residual encoders to extract multi-scale feature pyramids. At every decoder resolution, the corresponding spatial features are projected via $1\!{\times}\!1$ convolutions, concatenated, fused by a residual $3\!{\times}\!3$ block, and added to the main decoder stream. Together with the standard U-Net skip connections, this provides high-resolution structural guidance throughout the decoding process.


\subsection{Stage~2: One-Step Amortized Distillation}\label{sec:student}

The diffusion teacher learned in Stage~1 captures an empirical conditional distribution of high-resolution depth maps given the measurements and visual guidance. The teacher represents the learned conditional distribution $p_{\theta}(\mathbf{x}_0 \mid \mathbf{y})$.
Although the conditional teacher provides a strong reconstruction distribution, conditioning alone does not guarantee that a sample reproduces
the observed photon-arrival histogram when passed through the physical forward model. We therefore augment the learned conditional distribution with an
explicit sensor-domain consistency term. For a candidate depth map $\mathbf{x}_0$, assume $\mathcal{L}_{C}(\mathbf{x}_0; \mathbf{H})$
measure the discrepancy between the observed histogram $\mathbf{H}$ and the histogram rendered from $\mathbf{x}_0$ using the differentiable forward model described in Section~\ref{sec:consistency}. We then define the measurement-regularized target distribution as

\begin{equation}
    \widetilde{p}_{\theta}(\mathbf{x}_0 \mid \mathbf{y})
    \propto
    p_{\theta}(\mathbf{x}_0 \mid \mathbf{y})
    \exp\left[
        -\mathcal{L}_{C}(\mathbf{x}_0; \mathbf{H})
    \right],
    \label{eq:physics_regularized_target}
\end{equation}


We approximate this target with a one-step implicit distribution $q_{\phi}(\mathbf{x}_0 \mid \mathbf{y})$. The corresponding variational
objective is

\begin{equation}
    \phi^{\star} = \arg\min_{\phi} D_{\mathrm{KL}}
    \left[
        q_{\phi}(\mathbf{x}_0 \mid \mathbf{y})
        \,\|\, 
        \widetilde{p}_{\theta}(\mathbf{x}_0 \mid \mathbf{y})
    \right].
    \label{eq:variational_objective}
\end{equation}

We expand the KL divergence in~\eqref{eq:variational_objective} to obtain
\begin{equation}
    D_{\mathrm{KL}}
    \left[
        q_{\phi}
        \,\|\,
        \widetilde{p}_{\theta}
    \right]
    =
    \mathbb{E}_{q_{\phi}}
    \left[
        \log
        \frac{
            q_{\phi}(\mathbf{x}_0 \mid \mathbf{y})
        }{
            \widetilde{p}_{\theta}(\mathbf{x}_0 \mid \mathbf{y})
        }
    \right].
    \label{eq:kl_expansion}
\end{equation}

Substituting~\eqref{eq:physics_regularized_target} into~\eqref{eq:kl_expansion} and omitting constant when optimizing the student leads to the objective

\begin{equation}
    \mathcal{L}_{\mathrm{student}}
    =
    \mathbb{E}_{q_{\phi}}
    \left[
        \mathcal{L}_{C}(\mathbf{x}_0; \mathbf{H})
    \right]
    +
    D_{\mathrm{KL}}
    \left[
        q_{\phi}(\mathbf{x}_0 \mid \mathbf{y})
        \,\|\,
        p_{\theta}(\mathbf{x}_0 \mid \mathbf{y})
    \right].
    \label{eq:student_objective}
\end{equation}

The terms in Eq.~\eqref{eq:student_objective} encode the two requirements established from the outset: consistency with the physical observation model ($\mathcal{L}_{C}$, Sec.~\ref{sec:consistency}) and agreement of the student with the learned teacher conditional distribution ($D_{\mathrm{KL}}$, Sec.~\ref{sec:ikl}). This variational formulation provides a natural way to incorporate explicit data consistency for 1-step distillation.

{\br
\subsubsection{Enforcing data consistency on the estimated depth}
\label{sec:consistency}

Evaluating the data-fidelity term requires propagating the predicted high-resolution depth map back into the low-resolution histogram. Following Eq.~\eqref{eqt:Statistical_model_init}, we construct the unscaled signal response $u_{n,k}{=}c_n f(k-d_n)$, where $d_n$ is the predicted depth $\hat{\mathbf{x}}_0$ at pixel $n$ in time-bin units, and we set $c_n{=}\mathbf{R}_n$.
We assume a normalized Gaussian $f(\cdot)$ with $\sigma{=}2$ bins to approximate the impulse response of the STMicroelectronics sensor. The operator $\bsA$ pools responses over each low-resolution super-pixel footprint $\Omega_m$, giving $\bar{u}_{m,k}{=}(\bsA u_{:,k})_m{=}\sum_{n\in\Omega_m}u_{n,k}$.

Consistent with Eq.~\eqref{eqt:Gaussian_model}, the estimated histogram is $H^{\mathrm{est}}_{m,k}=g\bar{u}_{m,k}+b_m$, where $b_m$ is the low-resolution background level estimated as the median count per bin of the measured histogram. The gain $g$ matches
the total background-subtracted counts:
\begin{equation}
    g =
    \frac{\sum_{m,k} H_{m,k} - K\sum_m b_m}
         {\sum_{m,k} \bar{u}_{m,k}}.
\end{equation}
Under the Gaussian approximation, we use the mean squared error between the two histograms as the data-consistency loss:
\begin{equation}
    \mathcal{L}_{\mathrm{C}}
    = \frac{1}{MK}\sum_m\sum_k
    \left(H^{\mathrm{est}}_{m,k}-H_{m,k}\right)^2.
\end{equation}

}
\subsubsection{Prior Regularisation via Integral KL}\label{sec:ikl}

Minimising the KL-divergence term of~\eqref{eq:student_objective} directly in data space is unstable when $q_\phi$ and $p_\theta$ have misaligned supports~\cite{davi, diffinstruct}. We instead minimise the Integral KL (IKL), which compares the two across the diffusion trajectory,
\begin{equation}\label{eq:ikl}
  \mathcal{L}_{\mathrm{IKL}}(\phi)
  = \int_0^T w(t)\, D_{\mathrm{KL}}\!\bigl(q_\phi(\mathbf{x}_t\mid\mathbf{y})\,\big\|\,p_\theta(\mathbf{x}_t\mid\mathbf{y})\bigr)\,\mathrm{d}t ,
\end{equation}
{\br where $w(t)$ is a weighting coefficient defined as below,}

\begin{equation}
w(t) \;=\; \mathrm{SNR}_t^{-1} \;=\; \frac{\sqrt{1-\bar\alpha_t}}{\sqrt{\bar\alpha_t}},
\qquad \bar\alpha_t=\prod_{s\le t}(1-\beta_s),
\end{equation}

where $\beta_t$ is a diffusion noise schedule~\cite{ddpm}. The gradient of Eq.~\eqref{eq:ikl} is estimated by teacher-student score matching (see~\cite{davi, dmd-one-step-yin} for more details regarding the derivations),
\begin{equation}\label{eq:ikl_grad}
  \nabla_\phi\mathcal{L}_{\mathrm{IKL}}
  \approx \mathbb{E}_{q_\phi(\mathbf{x}_t|\mathbf{y}),\,t}
  \!\left[w(t)\bigl(\mathcal{S}_\psi(\mathbf{x}_t,t, \mathbf{y})-\mathcal{S}_\theta(\mathbf{x}_t,t, \mathbf{y})\bigr)\tfrac{\partial\mathbf{x}_t}{\partial\phi}\right] ,
\end{equation}
where $\mathcal{S}_\theta$ is the frozen teacher score and $\mathcal{S}_\psi$ is an
auxiliary score network tracking the evolving one-step generator, trained by denoising score matching,
\begin{equation}\label{eq:score_match}
  \mathcal{L}_S(\psi)
  = \mathbb{E}_{q_\phi(\mathbf{x}_t|\mathbf{y}),\,t}
    \bigl\lVert \mathcal{S}_\psi(\mathbf{x}_t,t, \mathbf{y})-\nabla_{\mathbf{x}_t}\log q_\phi(\mathbf{x}_t\mid\mathbf{y})\bigr\rVert_2^2 .
\end{equation}

Training alternates between updating $\mathcal{S}_\psi$ via Eq.~\eqref{eq:score_match} and updating the generator parameters $\phi$ through the combined objective 
$\nabla_{\phi}(w_{\mathrm{con}}\mathcal{L}_{C}+\mathcal{L}_{\mathrm{IKL}}+w_{\mathrm{rec}}\mathcal{L}_{\mathrm{rec}})$, where
$\mathcal{L}_{\mathrm{rec}} =\|\hat{\mathbf{x}}_0-\mathbf{x}_0\|_2^2$ is an auxiliary supervised reconstruction loss (see~\cite{davi} for more details).


\subsubsection{Perturbed Posterior Bridge}\label{sec:ppb}

Following~\cite{davi}, we adopt a Perturbed Posterior Bridge (PPB) to make the single-step generator robust across noise levels. {\color{black} By providing an intermediate set of trajectories, the PPB enables more guided training of the generator, $q_\phi$.} Since $\mathbf{H}$ and $\mathbf{x}_0$ live in different spaces, a coarse per-super-pixel depth $\tilde{y}_m$ is first obtained from the histogram using matched filtering with $\sigma{=}1$ bins. $\tilde{y}_m$ is then bilinearly upsampled to $H\!\times\!W$  to obtain $\tilde{\mathbf{y}}$. During training the generator input interpolates between this  $\tilde{\mathbf{y}}$ and the clean depth $\mathbf{x}_0$, perturbed by Gaussian noise,
\begin{equation}\label{eq:ppb}
  \mathbf{y}_a=(1-\sigma_a)\tilde{\mathbf{y}}+\sigma_a\mathbf{x}_0+h\,\Bar{\sigma}_a\mathbf{z},
  \quad \mathbf{z}\sim\mathcal{N}(\mathbf{0},\mathbf{I}),
\end{equation}

where $a {\in} [0,1], a\sim\operatorname{Beta}(3,1)$, and $\sigma_a {=} \frac{\int_a^1 \beta_t dt}{\int_0^1 \beta_t dt}$. We keep the perturbation $h=0.1$ unless otherwise stated, {\br and \(\bar{\sigma}_a = 1-\prod_{s=0}^{a}(1-\beta_s)\) is a monotonically increasing diffusion schedule.}


\subsection{Inference}\label{sec:inference}
At test time, the generator produces a reconstruction in a single forward pass,

\begin{equation}\label{eq:inference}
  \hat{\mathbf{x}}_0
  = q_\phi\!\bigl(\tilde{\mathbf{y}}+h\,\mathbf{z}, a=1, \;\mathbf{y}),
  \quad \mathbf{z}\sim\mathcal{N}(\mathbf{0},\mathbf{I}) .
\end{equation}

The data-consistency operator is used only in training and adds no inference cost; the inference pipeline is (i)~one frozen DepthAnything\,V2 pass for $\mathbf{d}_{\mathrm{DA}}$, (ii)~the upsampled coarse depth-map $\tilde{\mathbf{y}}$, and (iii)~one generator pass.

%% file: sections/04_experiments_v2.tex
\section{Experiments} \label{sec:experiments}

We evaluate the proposed method on both simulated and real single-photon data. We first describe the experimental setup
(Sec.~\ref{sec:exp-setup}), then report quantitative and qualitative results on simulated Middlebury scenes (Sec.~\ref{sec:exp-simulated-data}), validate on real captures acquired with an STMicroelectronics SPAD sensor (Sec.~\ref{sec:exp-real-data}), and analyze inference runtime (Sec.~\ref{sec:recon-time}).

\definecolor{rowlight}{RGB}{245, 247, 252}
\definecolor{rowdark}{RGB}{228, 233, 243}
\definecolor{avgrow}{RGB}{210, 220, 238}
\definecolor{avgcol}{RGB}{252, 241, 216}

\newcommand{\best}[1]{\textbf{#1}}
\newcommand{\scnd}[1]{\underline{#1}}
\newcommand{\avg}[1]{\cellcolor{avgcol}#1}

\begin{table*}[t]
\centering
\scriptsize
\renewcommand{\arraystretch}{1.18}
\setlength{\tabcolsep}{2.2pt}
\caption{Quantitative comparison on Middlebury scenes
($8\!\times\!8 \to 128\!\times\!128$) at two noise levels with a spacing of $7.65$\,mm/bin.
Each cell reports \textbf{RMSE} in mm\,/\,\textbf{mean surface normal error} in degrees (both $\downarrow$).
The clean configuration uses SBR\,=\,2.0, PPP\,=\,450 with clean
reflectivity. The noisy configuration uses SBR\,=\,1.0, PPP\,=\,300 with
degraded reflectivity (Poisson $\gamma{=}100$, Gaussian $\sigma_r{=}0.02$).
DA: DepthAnything\,V2-base aligned to the measured histograms,
DELTAR~\cite{deltar}, Marigold-DC~\cite{marigold-dc}, and
DVSR~\cite{sun2023consistent}.
Within each scene, \best{bold} marks the best (lowest) value and
\scnd{underline} the second best, applied to each metric independently.
The shaded rightmost column reports the mean over all twelve scenes.}
\label{tab:quant-results}

\resizebox{\textwidth}{!}{%
\begin{tabular}{lccccccccccccc}
\toprule
\textbf{Method}
& \textbf{Art}
& \textbf{Baby}
& \textbf{Books}
& \textbf{Bowling}
& \textbf{Cloth}
& \textbf{Dolls}
& \textbf{Lampshade}
& \textbf{Laundry}
& \textbf{Moebius}
& \textbf{Monopoly}
& \textbf{Plastic}
& \textbf{Reindeer}
& \avg{\textbf{Avg.}} \\
\midrule

\multicolumn{14}{c}{
\textbf{SBR\,=\,2.0,\; PPP\,=\,450 \;(Clean Reflectivity)}
} \\
\midrule

\rowcolor{rowlight}
DA (aligned)
& 41.7/26.7 & 10.8/\best{11.3} & 25.3/24.6
& 57.7/\scnd{18.0} & 13.9/\scnd{15.1} & 29.3/\best{27.1}
& 39.6/19.0 & 54.0/\best{22.2} & 30.9/23.3
& 43.5/17.0 & 22.5/\scnd{12.1} & 27.1/\best{17.7}
& \avg{33.0/\scnd{19.5}} \\

\rowcolor{rowdark}
DELTAR
& 73.4/64.0 & 28.6/51.6 & 30.7/51.7
& 63.2/50.5 & 20.8/45.5 & 26.0/52.7
& 52.6/48.4 & 53.8/54.4 & 35.0/56.7
& 74.5/58.3 & 24.5/41.6 & 39.6/49.5
& \avg{43.6/52.1} \\

\rowcolor{rowlight}
Marigold-DC
& 68.9/55.8 & 20.8/19.8 & 30.6/37.5
& 47.3/35.7 & 8.4/31.4 & 20.6/40.4
& 34.1/34.7 & 53.2/55.1 & 30.4/38.4
& 33.8/26.0 & 16.0/23.1 & 31.1/41.1
& \avg{32.9/36.6} \\

\rowcolor{rowdark}
DVSR
& 74.5/51.4 & 24.9/29.8 & 39.9/39.9
& 52.7/36.4 & 12.5/22.5 & 24.9/44.7
& 63.5/39.0 & 46.3/41.3 & 39.7/38.3
& 46.0/39.2 & 20.7/28.9 & 42.3/34.0
& \avg{40.7/37.1} \\

\rowcolor{rowlight}
D2ToF (teacher)
& \scnd{24.6}/\best{23.3} & \scnd{8.9}/\scnd{13.1} & \scnd{11.2}/\best{19.9}
& \scnd{16.6}/\best{16.2} & \scnd{5.9}/\best{14.6} & \scnd{12.9}/27.8
& \scnd{20.3}/\best{14.8} & \scnd{17.3}/\scnd{22.4} & \scnd{14.8}/\best{22.9}
& \scnd{14.7}/\best{12.9} & \scnd{6.0}/\best{11.1} & \scnd{13.1}/\scnd{18.0}
& \avg{\scnd{13.9}/\best{18.1}} \\

\rowcolor{avgrow}
\textbf{D2ToF}
& \best{21.2}/\scnd{25.2} & \best{7.1}/16.0 & \best{9.7}/\scnd{21.7}
& \best{14.9}/18.7 & \best{4.1}/\best{14.6} & \best{11.0}/\scnd{27.2}
& \best{17.4}/\scnd{18.4} & \best{13.6}/24.2 & \best{12.5}/\scnd{23.2}
& \best{12.1}/\scnd{15.5} & \best{4.7}/15.4 & \best{12.0}/20.5
& \avg{\best{11.7}/20.1} \\

\midrule
\multicolumn{14}{c}{
\textbf{SBR\,=\,1.0,\; PPP\,=\,300 \;(Noisy Reflectivity)}
} \\
\midrule

\rowcolor{rowlight}
DA (aligned)
& 63.9/\scnd{38.9} & 34.8/\best{22.8} & 81.3/31.4
& 78.2/29.6 & 24.3/\best{23.7} & 46.8/\scnd{34.7}
& 57.3/\best{27.5} & 66.5/36.5 & 59.7/\best{32.5}
& 146.1/\scnd{23.7} & 40.9/23.6 & 33.3/\scnd{22.9}
& \avg{61.1/29.0} \\

\rowcolor{rowdark}
DELTAR
& 70.6/64.2 & 28.5/51.5 & 30.7/51.4
& 60.6/50.2 & 21.5/45.5 & 26.1/52.6
& 55.4/48.1 & 53.9/54.4 & 34.8/56.6
& 74.5/58.5 & 24.8/41.7 & 38.5/49.6
& \avg{43.3/52.0} \\

\rowcolor{rowlight}
Marigold-DC
& 77.8/70.1 & 40.3/46.4 & 40.3/56.0
& 60.1/46.9 & 16.0/52.8 & 25.8/47.1
& 70.8/69.2 & 51.9/61.8 & 41.4/51.6
& 48.5/45.8 & 26.1/40.5 & 51.7/43.7
& \avg{45.9/52.7} \\

\rowcolor{rowdark}
DVSR
& 73.6/51.3 & 37.1/45.2 & 38.9/39.1
& 55.5/36.2 & 14.1/24.0 & 27.3/46.3
& 56.6/41.7 & 41.0/43.4 & 37.0/39.6
& 84.4/42.9 & 23.4/31.0 & 40.2/36.2
& \avg{44.1/39.7} \\

\rowcolor{rowlight}
D2ToF (teacher)
& \scnd{48.3}/39.5 & \scnd{19.4}/\scnd{23.6} & \scnd{18.4}/\scnd{26.9}
& \scnd{31.9}/\scnd{27.7} & \scnd{10.2}/\best{23.7} & \scnd{17.8}/37.1
& \scnd{42.5}/\scnd{29.4} & \scnd{27.6}/\scnd{35.2} & \scnd{26.5}/\scnd{33.6}
& \scnd{34.8}/\best{23.4} & \scnd{11.5}/\best{18.5} & \scnd{20.3}/\best{21.9}
& \avg{\scnd{25.8}/\scnd{28.4}} \\

\rowcolor{avgrow}
\textbf{D2ToF}
& \best{40.5}/\best{38.3} & \best{15.4}/\scnd{23.6} & \best{15.5}/\best{26.4}
& \best{28.2}/\best{26.4} & \best{7.9}/\scnd{23.8} & \best{14.6}/\best{33.6}
& \best{39.5}/29.5 & \best{21.7}/\best{31.9} & \best{22.7}/33.9
& \best{28.4}/24.0 & \best{9.4}/\scnd{21.0} & \best{17.2}/23.2
& \avg{\best{21.7}/\best{28.0}} \\

\bottomrule
\end{tabular}%
}
\end{table*}
\subsection{Experimental Setup} \label{sec:exp-setup}

\paragraph{Training data and simulation} \label{par:exp-training-data-and-simulation}

The teacher and student networks are trained on depth-reflectivity pairs drawn from the Hypersim~\cite{Hypersim} and MPI Sintel~\cite{MPI} datasets for all the simulated and real data evaluation. From each sequence we extract $128\times128$ patches and synthesise the corresponding low-resolution ToF histograms following the observation model in Eq.~\eqref{eqt:Statistical_model_init}. The histogram is computed on a $16\times$ spatially downsampled grid, yielding an $8\times8$ spatial resolution with $K=100$ time bins per pixel. Photon arrivals are drawn as Poisson realizations, with a fixed signal-to-background ratio (SBR) at $\text{SBR}{=}2$ and an impulse response of $\sigma=2$ bins. For the Hypersim dataset (43140 frames), the photons-per-pixel (PPP) is kept fixed at $\text{PPP}{=}450$, whereas the MPI dataset (2128 frames) uses a variable PPP in the range $[550, 800]$. We apply geometric augmentation (horizontal flips and $90^{\circ}$ rotations) jointly to the depth, reflectivity, and histogram tensors. To account for the imperfect co-registration between the SPAD array and the guide camera, we additionally apply a random misalignment to the histogram during training. With probability $0.5$, a random translation and scaling is applied to the $8\times8$ grid about its center. The horizontal and vertical offsets are drawn independently from $\mathcal{U}(-10, 10)$ image pixels (equivalent to $\pm0.625$ of a cell as one cell spans $16$ px on $128\times128$ image) and the scale factor from $\mathcal{U}(0.92, 1.08)$.

\paragraph{Implementation details}
Our U-Net backbone is trained using the $\epsilon$-prediction objective in Eq.~\eqref{eq:teacher_loss} with $T=1000$ diffusion timesteps. The teacher is trained for $66$k iterations using Adam with a constant learning rate of $10^{-4}$ and a batch size of $48$. The student is trained for $60$k iterations with a batch size of $6$. For the student, we use AdamW with a constant learning rate of $10^{-4}$ and a weight decay of $0.01$. Training is performed using three NVIDIA RTX~3090 GPUs for $14$ hours. In all experiments, we set $w_{\mathrm{con}}{=}12$ and $w_{\mathrm{rec}}{=}3$. The DepthAnything~V2 model  (ViT-B backbone) remains frozen throughout both training stages. At inference, the teacher uses $250$ DDPM sampling steps, whereas the proposed student produces a reconstruction in a single forward pass. All inference-time experiments are performed on a single NVIDIA RTX~3090 GPU with a batch size of $1$.


\subsection{Evaluation metrics and baseline methods} \label{subsec:eval-metrics-baseline}

In simulated case with ground truth depth, we quantitatively evaluate performance using the root mean-squared error (RMSE) (in mm) and the mean surface normal error (SNE) (in degrees).  For real data, ground truth depth is not available and we use {\br histogram consistency} to evaluate performance.  
We compare our method against a recent diffusion-based depth completion work Marigold-DC~\cite{marigold-dc}, a single-photon depth super-resolution methods, DVSR~\cite{sun2023consistent} and DELTAR~\cite{deltar}. Finally, we include our multi-step teacher to quantify the fidelity against the one-step student.

\begin{figure*}
    \centering
    \includegraphics[width=0.9\linewidth]{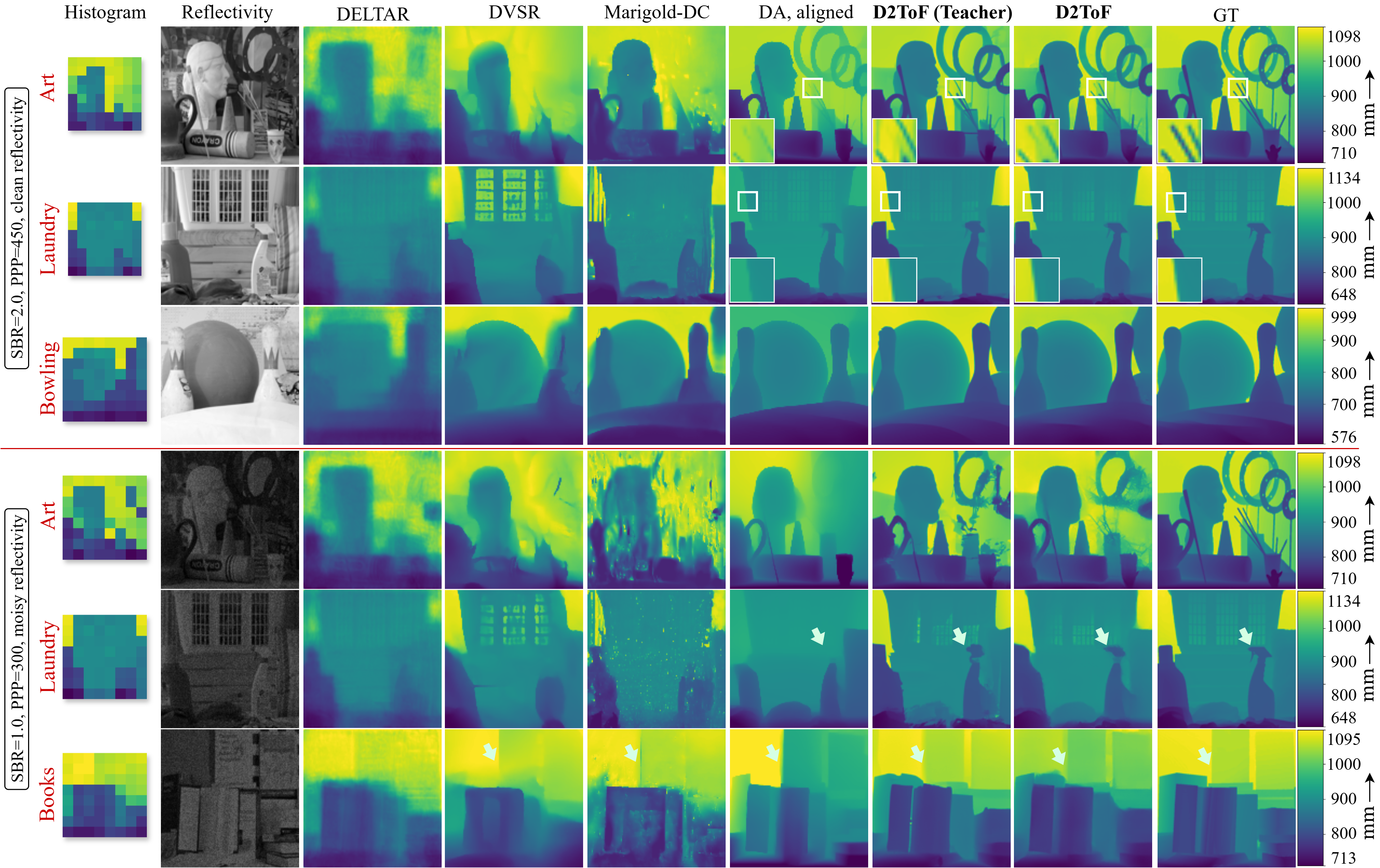}
    \caption{(Top) Comparison with different methods when both histograms (SBR=2.0, PPP=450.0) and reflectivity maps are in clean scenario. Our method achieves superior results compared to other reconstruction methods. (Bottom) Qualitative comparison of different methods under noisy conditions. The inputs are degraded by both reflectivity noise (Poisson $\gamma{=}100$, Gaussian $\sigma_r{=}0.02$) and ToF histogram noise (SBR\,=\,1.0, PPP\,=\,300). Our proposed approach significantly outperforms existing baselines under such noisy scenarios.}
    \label{fig:simulated-results}
\end{figure*}



\textbf{DepthAnything (DA).} We apply DepthAnything\,V2 (ViT-B)~\cite{depthanythingv2} directly to the high-resolution reflectivity image. Since DA predicts relative disparity, we align it to our measurement histogram. Matched filtering ($\sigma{=}3$ bins) of the measured histogram gives a coarse metric depth $\mathbf{y}_{\text{hist}} \in \mathbb{R}^{8 \times 8}$, and the DA output is average-pooled onto the same grid to give $\bar{x}_{\text{DA}}$. We regress inverse depth,

\begin{equation}
    \frac{1}{Z_m} \;\approx\; a\,\bar{x}_{\text{DA},m} + b ,
    \qquad m = 1,\dots,M ,
\end{equation}

solve for $(a,b)$ by least squares over the $M = 64$ super-pixels, and invert the fitted map. Coefficients are re-estimated for every scene and noise level.

\textbf{DELTAR} \cite{deltar} predicts dense metric depth from a lightweight $8\times8$ zone ToF sensor paired with an RGB image, matching the input/output shapes of this benchmark. It does not consume the waveform, instead each zone is summarized by a Gaussian having $(\mu,\sigma)$ pair.

\textbf{Marigold-DC}~\cite{marigold-dc} performs monocular depth completion using an image together with sparse metric depth measurements. To adapt our $8\times8$ ToF measurements to this input format, we first estimate one metric depth value for each ToF zone using matched filtering. Each estimated zone depth is then placed at the centre of its corresponding $16\times16$ region on a $128\times128$ sparse-depth map. This produces $64$ sparse depth anchors, while all unmeasured pixels are set to zero to indicate missing depth. The resulting sparse-depth map and the high-resolution guidance image are provided to Marigold-DC. We use the released depth checkpoint with trailing timestep spacing and run inference for $50$ DDIM steps.

\textbf{DVSR}~\cite{sun2023consistent} is a video-based super-resolution network that consumes a per-pixel peak depth rather than a waveform along with an RGB image. We report the dynamic regime it was designed for: the argmax of the measured histogram is bicubic-upsampled to guidance resolution, panned one pixel per frame across a ten-frame clip, finally metrics calculation uses the zero-offset reference frame. 

\subsection{Results on simulated data} \label{sec:exp-simulated-data}

\paragraph{Testing data}
We simulate high-resolution SPAD histograms with a single depth return per pixel and evaluate on $12$ scenes from the Middlebury stereo dataset~\cite{Middlebury} (Art, Bowling, Laundry, Reindeer, Moebius, Lampshade, Dolls, Baby, Books, Monopoly, Cloth, and Plastic), which provide aligned high-resolution reflectivity and depth maps. Missing depth values arising from stereo occlusions are filled using median filtering prior to histogram simulation. The HR histograms are used to build LR versions that have $100$ time bins at an $8\times8$ spatial resolution, with $128\times128$ reflectivity maps. We consider two operating points: clean settings (SBR$=2.0$, PPP$=450$) with clean reflectivity, and a challenging dark condition (SBR$=1.0$, PPP$=300$) where the guidance reflectivity is degraded by Poisson and Gaussian noise ($\gamma=100$, $\sigma_r=0.02$).



\paragraph{Quantitative results}
{\br Table~\ref{tab:quant-results} reports RMSE (in mm) and SNE (in degrees), averaged across $12$ scenes under clean and noisy conditions. For each method, scene, and condition, we generate $50$ reconstructions from the same fixed input measurement, compute the metrics of each against ground truth, and average the $50$ values.}
Our proposed one-step model substantially outperforms all baselines in depth accuracy in both scenarios. Under clean conditions, it achieves an average RMSE of $11.7$~mm, roughly one-third the error of the recent Marigold-DC ($32.9$~mm) and a significant improvement over the histogram-aligned DA baseline ($33.0$~mm). The one-step student model surpasses its multi-step teacher ($13.9$~mm) on every scene.
{\br 
SNE shows a slightly different trend from depth RMSE. Under clean conditions, the teacher achieves a lower angular error than the one-step student ($18.1^\circ$ versus $20.1^\circ$), despite their similar RMSE. Under noisy conditions, however, the student achieves the lowest value in both RMSE ($21.7$~mm) and SNE ($28.0^\circ$). DA also performs well in terms of surface normals under clean conditions ($19.5^\circ$), despite having a considerably higher RMSE ($33.0$~mm versus $11.7$~mm for D2ToF). This difference arises because SNE measures local surface orientation rather than absolute depth accuracy. Furthermore, D2ToF retains the lowest mean RMSE under the tested degradation.} While noise propagates visibly into the baseline estimates and severely degrades the intermediate DA predictions, our approach consistently recovers coherent depth structures. Crucially, we achieve this robustness zero-shot, requiring no retraining or fine-tuning on noisy reflectivity. We note finally that the analysis of the  histogram consistency is grouped in Sec.~\ref{subsec:ablation-con-loss}.


\paragraph{Qualitative results}
Figure~\ref{fig:simulated-results} presents qualitative comparisons in both clean and noisy settings. The baseline methods, DELTAR~\cite{deltar} and DVSR~\cite{sun2023consistent}, fail to recover sharp depth discontinuities from the $8\times8$ inputs, yielding over-smoothed reconstructions. Although Marigold-DC~\cite{marigold-dc} performs relatively well in uniform scenes like \textit{Bowling}, it generates subpar results in scenes containing intricate details, such as \textit{Art}. The performance degrades further in the noisy setting. The reflectivity-only DA model (aligned to measurement histogram, explained in Sec.~\ref{subsec:eval-metrics-baseline}) captures plausible scene structures but fails severely in noisy scenarios. Conversely, our one-step framework successfully reconstructs the global geometry and fine structural details in both environments.
\subsection{Results on real data} \label{sec:exp-real-data}
To evaluate our approach in practical scenarios, we utilize the VL53L8 time-of-flight (ToF) sensor manufactured by STMicroelectronics. This device actively illuminates the scene with 940-nm light pulses, employing single-photon avalanche diode (SPAD) detectors and dedicated timing electronics to capture the time of arrival of reflected photons, thereby generating histograms of photon counts.

For our real-world validation, the ToF histograms are captured with the sensor configured to an $8\times8$ spatial resolution, yielding $144$ time bins per pixel. We isolate the first $100$ bins for processing, discarding the trailing background data. {\br A Kinect sensor (positioned beside)} provides the high-resolution reflectivity maps used for guidance, together with a reference high-resolution depth map for visual comparison. We collect scenes under both well-lit and low-light conditions to characterise behaviour across different photon regimes.

\begin{figure*}
    \centering
    \includegraphics[width=0.87\linewidth]{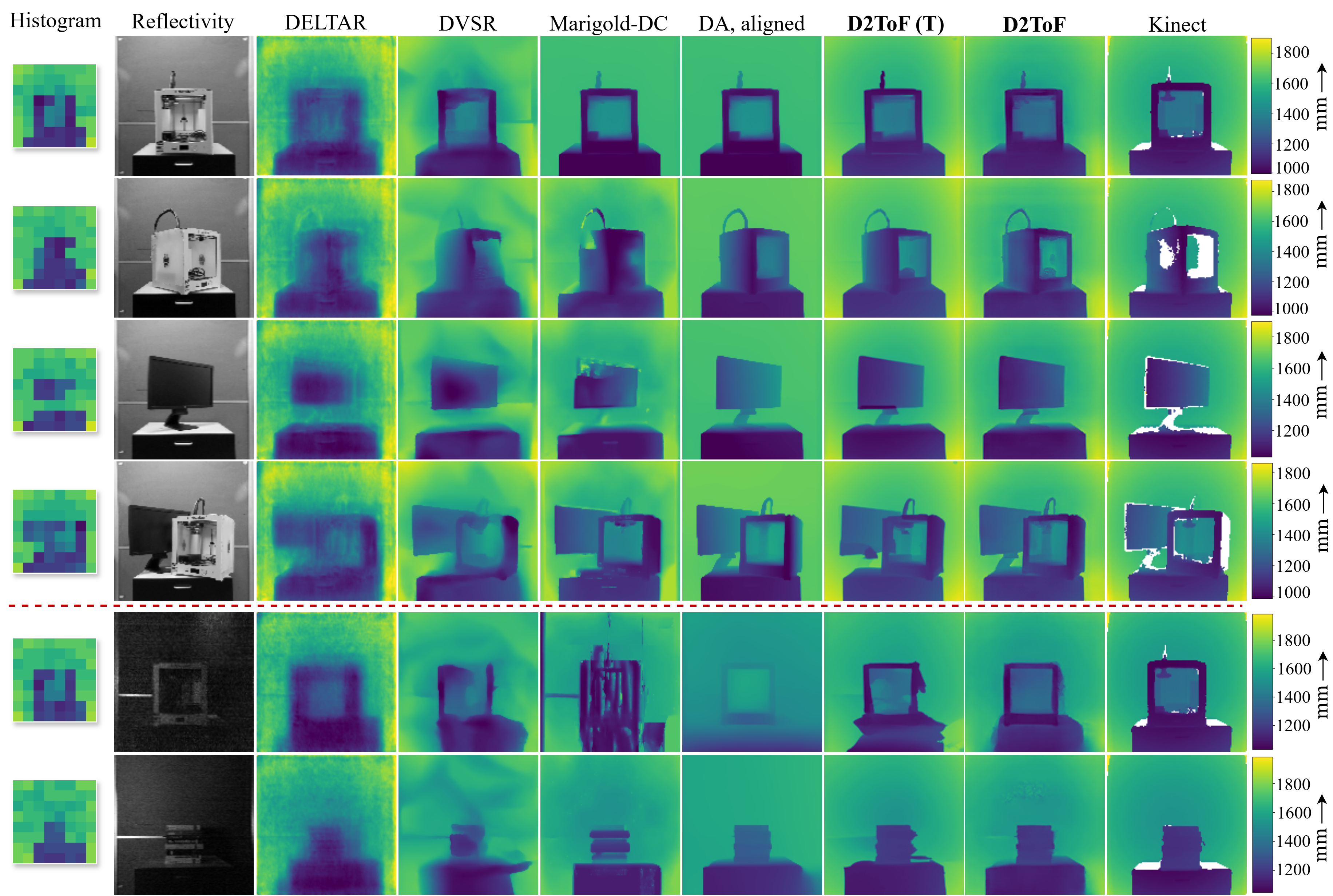}
    \caption{Comparison with different methods for real data collected using an inexpensive STMicroelectronics sensor in $8\times8$ mode. The reflectivity maps ($128\times128$) are provided by a Kinect sensor connected adjacent to the SPAD.}
    \label{fig:real-results-clean}
\end{figure*}

Figure~\ref{fig:real-results-clean} shows reconstructions under well-lit conditions (first four rows) and low-light conditions (last two rows). Despite being trained purely on simulated data, the qualitative results demonstrate transfer to real sensor measurements and produces sharp, geometrically consistent depth maps, recovering object boundaries that are absent from the raw $8\times8$ waveforms.
Reflectivity-based algorithms, such as DA and Marigold-DC, perform well when clean reflectivity data are available. However, their performance degrades under low-light conditions due to reduced photon counts, which make reconstruction more challenging. In contrast, the proposed algorithm demonstrates greater robustness under these conditions.
Figure~\ref{fig:real-plot} displays the measured waveforms for nine pixels in the central region of the sensor, illustrating the photon-arrival statistics underlying the reconstruction.

\subsection{Reconstruction time} \label{sec:recon-time}

A central advantage of the proposed framework is its inference speed. Although the method ingests the full ToF histogram, the amortized single-step formulation enables reconstruction in a single forward pass. As summarized in Figure~\ref{fig:speed-quality}, our one-step method reconstructs a scene in approximately $83$\,\textit{ms} on average, of which roughly $39$\,\textit{ms} is attributable to the frozen DA. {\br For DVSR, Figure~\ref{fig:speed-quality} reports the average inference time per frame over the 10-frame clip}. The multi-step teacher requires $11$\,\textit{s} per sample owing to its $250$-step DDPM sampling process. The one-step student is therefore approximately $130\times$ faster than the teacher while being slightly better in average RMSE.  Even without DA, our model performs relatively well at a faster inference speed of $41$\,\textit{ms} (see Figure~\ref{fig:speed-quality}). This places our framework in a highly favorable region of the speed-quality trade-off, putting it well within reach of interactive frame rates on a single RTX 3090 GPU. Furthermore, the differentiable data-consistency term is utilized exclusively during training, thereby incurring no additional inference-time overhead.



\section{Ablation Study}\label{sec:exp-ablation}

The ablations below are evaluated under 4 operating points that degrade the two input modalities (Table.~\ref{tab:noise-settings}). Quantitative results are averaged over the $12$ Middlebury scenes and reported as RMSE ({mm}). For selected experiments, we also report the mean RMSE and standard deviation ($\pm$) computed from $50$ stochastic reconstructions of the same input.

\begin{table}[!htb]
\centering
\caption{Operating points used throughout the ablation study.}
\label{tab:noise-settings}
\begin{tabular}{lccccc}
\toprule
& \multicolumn{2}{c}{Histogram} & \multicolumn{2}{c}{Reflectivity} \\
\cmidrule(lr){2-3}\cmidrule(lr){4-5}
Setting & SBR & PPP & $\gamma$ & $\sigma_r$ \\
\midrule
\textit{Clean}   & 2.0 & 450 & --  & --   \\
\textit{Noise~1} & 2.0 & 450 & 100 & 0.02 \\
\textit{Noise~2} & 1.0 & 300 & 100 & 0.02 \\
\textit{Noise~3} & 1.0 & 300 & 40  & 0.05 \\
\bottomrule
\end{tabular}
\end{table}
\subsection{Effect of the visual conditionings}\label{subsec:ablation-conditionings}
\begin{figure}[!htb]
    \centering
    \includegraphics[width=1\linewidth]{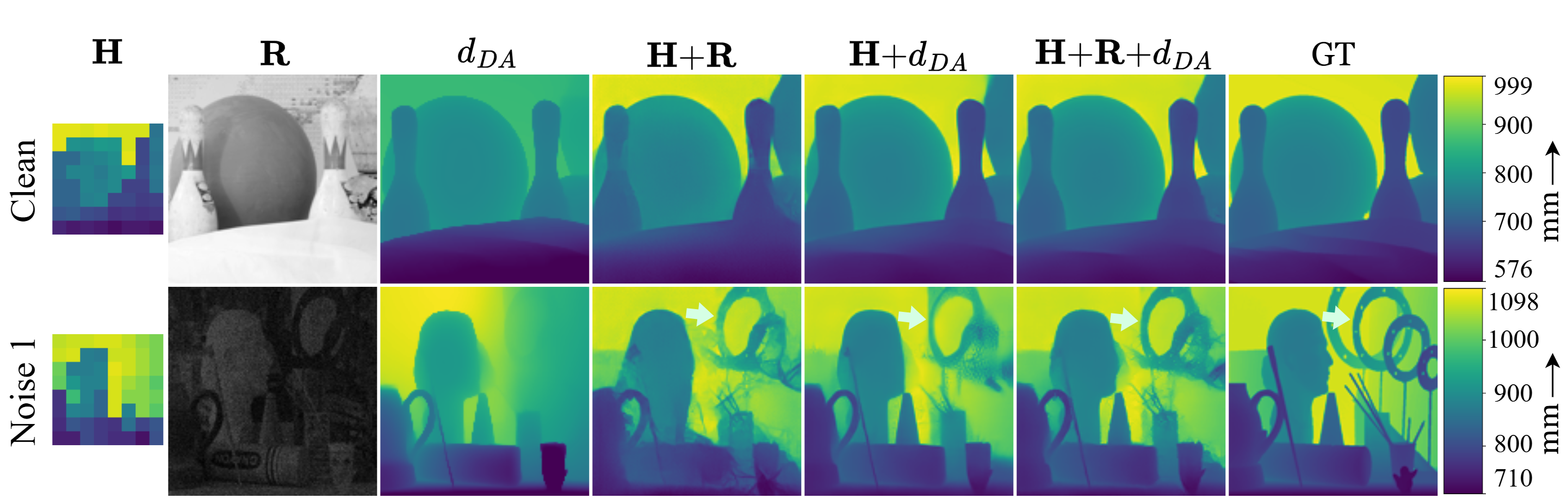}
    \caption{Effect of reflectivity map $\mathbf{R}$ and DA output $\mathbf{d}_{\mathrm{DA}}$ when either of them is used as a condition on the network decoder of D2ToF.}
    \label{fig:ablation-conditioning}
\end{figure}

We ablate the two visual conditioning signals: the high-resolution reflectivity map $\mathbf{R}$ and the DA output $\mathbf{d}_{\mathrm{DA}}$ to quantify their individual contributions and, critically, the robustness of D2ToF when the reflectivity itself is degraded.
We compare the full model against two reduced variants: one without DA ($\mathbf{H}{+}\mathbf{R}$) and one without the reflectivity channel ($\mathbf{H}{+}\mathbf{d}_{\mathrm{DA}}$) and, as a reference, the monocular prediction alone. Quantitative results are reported in Table~\ref{tab:ablation-conditioning} and visual results are shown in Figure~\ref{fig:ablation-conditioning}.

\definecolor{rowlight}{RGB}{245, 247, 252}
\definecolor{rowdark}{RGB}{228, 233, 243}
\definecolor{fullrow}{RGB}{210, 220, 238}

\begin{table}[!htb]
\centering
\footnotesize
\renewcommand{\arraystretch}{1.25}
\setlength{\tabcolsep}{3pt}
\caption{Ablation of the reflectivity ($\mathbf{R}$) and DA-output ($\mathbf{d}_{\mathrm{DA}}$) conditioning. The highlighted row is the full model.}

\label{tab:ablation-conditioning}
\begin{tabular}{l|ccc}
\toprule
\textbf{Configuration}  & \textbf{\textit{Clean}} & \textbf{\textit{Noise~1}} & \textbf{\textit{Noise~2}}\\
\midrule
\rowcolor{rowlight} Monocular ($\mathbf{d}_{\mathrm{DA}}$ only)                                                 & 33.0      & 61.2      & 61.1 \\
\rowcolor{rowdark}  with $\mathbf{R}$ only ($\mathbf{H}{+}\mathbf{R}$)                                          & 18.19 $\pm$ 0.14      & 25.83 $\pm$ 0.08      & 26.69 $\pm$ 0.05\\
\rowcolor{rowlight} with $\mathbf{d}_{\mathrm{DA}}$ only ($\mathbf{H}{+}\mathbf{d}_{\mathrm{DA}}$)              & 12.59 $\pm$ 0.06      & 21.69 $\pm$ 0.07      & 23.11 $\pm$ 0.07\\
\rowcolor{fullrow} \textbf{Full ($\mathbf{H}{+}\mathbf{R}{+}\mathbf{d}_{\mathrm{DA}}$)}                         & \textbf{11.71 $\pm$ 0.06}      & \textbf{20.16 $\pm$ 0.05}      & \textbf{21.75 $\pm$ 0.05} \\
\bottomrule
\end{tabular}
\end{table}

Two findings emerge. First, the histogram adds robustness to noise. The aligned monocular baseline, which sees no range information,
degrades sharply as the reflectivity is corrupted, since its RMSE rising from $33.0$ to $61.2$~mm, whereas every variant that
includes the histogram keeps RMSE below $30.0$~mm across all three conditions. Under noisy conditions, the monocular error is nearly $3$ times that of the full model.

Second, the two visual conditionings are complementary and provide robustness. On all scenarios, the full
$\mathbf{H}{+}\mathbf{d}_{\mathrm{DA}}{+}\mathbf{R}$ variant has the lowest RMSE. Under noise, the full model retains the lowest RMSE ($20.16$ and $21.75$). The results confirm that both signals are needed for the best overall trade-off. Removing either $\mathbf{d}_\mathrm{DA}$ or $\mathbf{R}$ produces more error under noisy conditions compared to the full variant.


\subsection{Effect of $\mathcal{L}_C$}
\label{subsec:ablation-con-loss}

The data-consistency term $\mathcal{L}_C$ is applied only during training. To verify that the reconstruction agrees with the measurement at test time, we take the depth map produced by each method, re-render it into a low-resolution histogram, and compare it with the measured histogram at \textit{clean} case.
Tables~\ref{tab:ablation-con-loss} and~\ref{tab:ablation-histogram-comparison} show the effect of the consistency loss in our framework. Including $\mathcal{L}_C$ reduces mean depth RMSE across all four simulated conditions. When only the reflectivity guidance is degraded (\textit{Noise~1}), RMSE decreases from $21.26$ to $20.16$~mm using $\mathcal{L}_C$, although training uses only clean reflectivity images. The loss also reduces the mean squared error ($\|\mathbf{H}_{\mathrm{est}}-\mathbf{H}\|_2^2$) from $8.62$ to $6.74$ on the 12 simulated Middlebury scenes under the clean setting and from $19.70$ to $16.12$ on the four well-lit real captures.

\definecolor{rowlight}{RGB}{245, 247, 252}
\definecolor{rowdark}{RGB}{228, 233, 243}
\definecolor{fullrow}{RGB}{210, 220, 238}
\begin{table}[!htb]
\centering
\footnotesize
\renewcommand{\arraystretch}{1.25}
\setlength{\tabcolsep}{3pt}
\caption{Ablation of the consistency loss $\mathcal{L}_C$ under noise levels defined in Table~\ref{tab:noise-settings}.}

\label{tab:ablation-con-loss}
\begin{tabular}{l|cccc}
\toprule
\textbf{Config.}                              & \textbf{\textit{Clean}}    & \textbf{\textit{Noise~1}}  & \textbf{\textit{Noise~2}}  & \textbf{\textit{Noise~3}} \\
\midrule
\rowcolor{rowlight} w/o $\mathcal{L}_C$             & 12.16 $\pm$ 0.13              & 21.26 $\pm$ 0.09              & 22.42 $\pm$ 0.08             & 25.98 $\pm$ 0.08 \\

\rowcolor{fullrow} \textbf{with $\mathcal{L}_C$}    & \textbf{11.71 $\pm$ 0.06}     & \textbf{20.16 $\pm$ 0.05}     & \textbf{21.75 $\pm$ 0.05}              & \textbf{25.01 $\pm$ 0.06} \\
\bottomrule
\end{tabular}
\end{table}
\begin{table}[!htb]
\centering
\caption{{\br Histogram reconstruction error ($\|\mathbf{H}_{\mathrm{est}}-\mathbf{H}\|_2^2$; lower is better), averaged over 12 \textit{clean} simulated Middlebury scenes and four real scenes (well-lit scenes from Figure~\ref{fig:real-results-clean}). GT is unavailable for the real scenes.}}
\label{tab:ablation-histogram-comparison}
\setlength{\tabcolsep}{2.75pt}
\begin{tabular}{ccccccc}
\toprule
\thead{\textbf{DELTAR}} &
\thead{\textbf{Marigold}\\\textbf{DC}} &
\thead{\textbf{DVSR}} &
\thead{\textbf{D2ToF}\\\textbf{(Teacher)}} &
\thead{\textbf{D2ToF}\\\textbf{(w/o Con. Loss)}} &
\thead{\textbf{D2ToF}} &
\thead{\textbf{GT}} \\
\midrule
\multicolumn{7}{c}{\emph{Simulated Middlebury (12-scene mean)}} \\
37.04 & 13.04 & 25.28 & 12.91 & 8.62 & 6.74 & 4.75 \\
\midrule
\multicolumn{7}{c}{\emph{Real data (4-scene mean)}} \\
53.41 & 18.00 & 30.02 & 21.04 & 19.70 & 16.12 & -- \\
\bottomrule
\end{tabular}
\end{table}

\begin{figure}[!htb]
    \centering
    \includegraphics[width=1\linewidth]{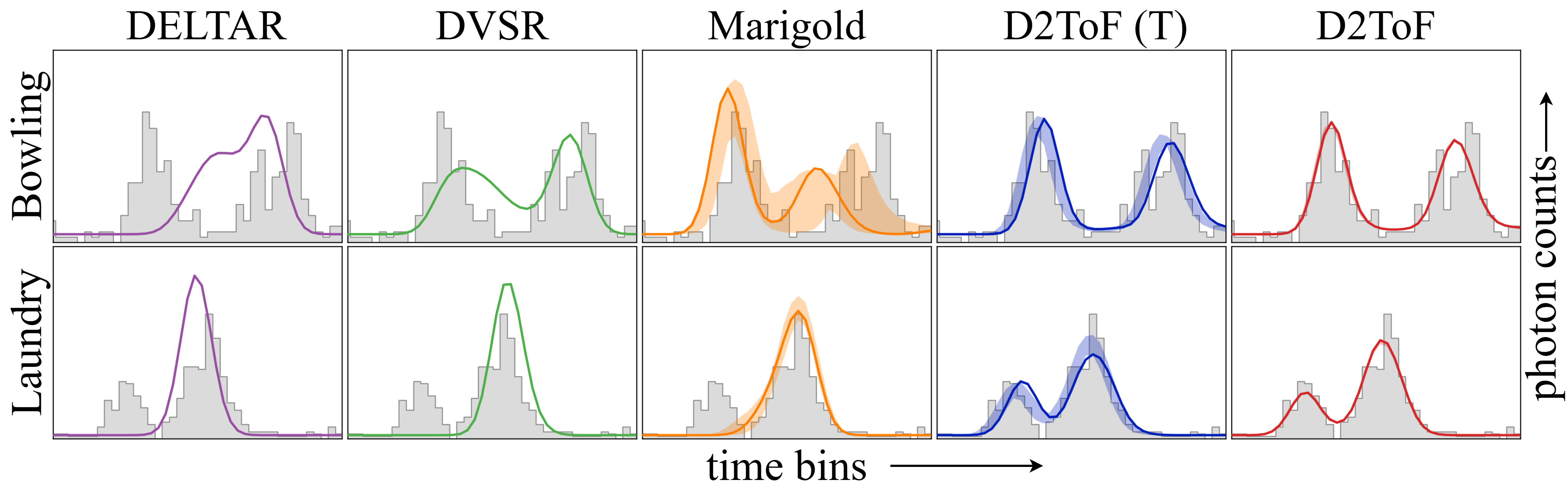}
    \caption{Histogram consistency for one low-resolution pixel from the \textit{Bowling} (top) and \textit{Laundry} (bottom) scenes. The gray bars show the measured
    ToF histogram, while the colored curves show the histograms re-rendered from the depth estimates produced by each method. We generate 50 samples for each method to generate the mean (solid line) and the standard deviation (shaded area). The proposed methods (D2ToF-Teacher and D2ToF) reproduce the measured temporal structure more closely than the compared methods.}
    \label{fig:histogram-consistency}
\end{figure}

Figure~\ref{fig:histogram-consistency} shows this comparison for a representative low-resolution pixel from the \textit{Laundry} and \textit{Bowling} scenes. Both our teacher and our student place the reconstructed peak at the measured depth and reproduce its shape, whereas the baselines deviate from the measured waveform.

\subsection{Effect of DepthAnything Model Size}
\label{subsec:ablation-depth-estimators}

We also evaluate the smaller DA variant to examine the accuracy-runtime trade-off of the monocular guidance module. Replacing
\textit{DA-base} with \textit{DA-small} reduces the reconstruction time from $83$\,\textit{ms} to $69$\,\textit{ms}, while increasing the average RMSE from $11.7$ to $13.1$ in the \emph{clean} setting and from $20.2$ to $22.2$ under \emph{Noise~1}. This provides a lower-latency alternative when a modest reduction
in reconstruction accuracy is acceptable.

\subsection{Uncertainty Estimation} \label{subsec:ablation-uncertainty}

\begin{figure}[!htb]
    \centering
    \includegraphics[width=1\linewidth]{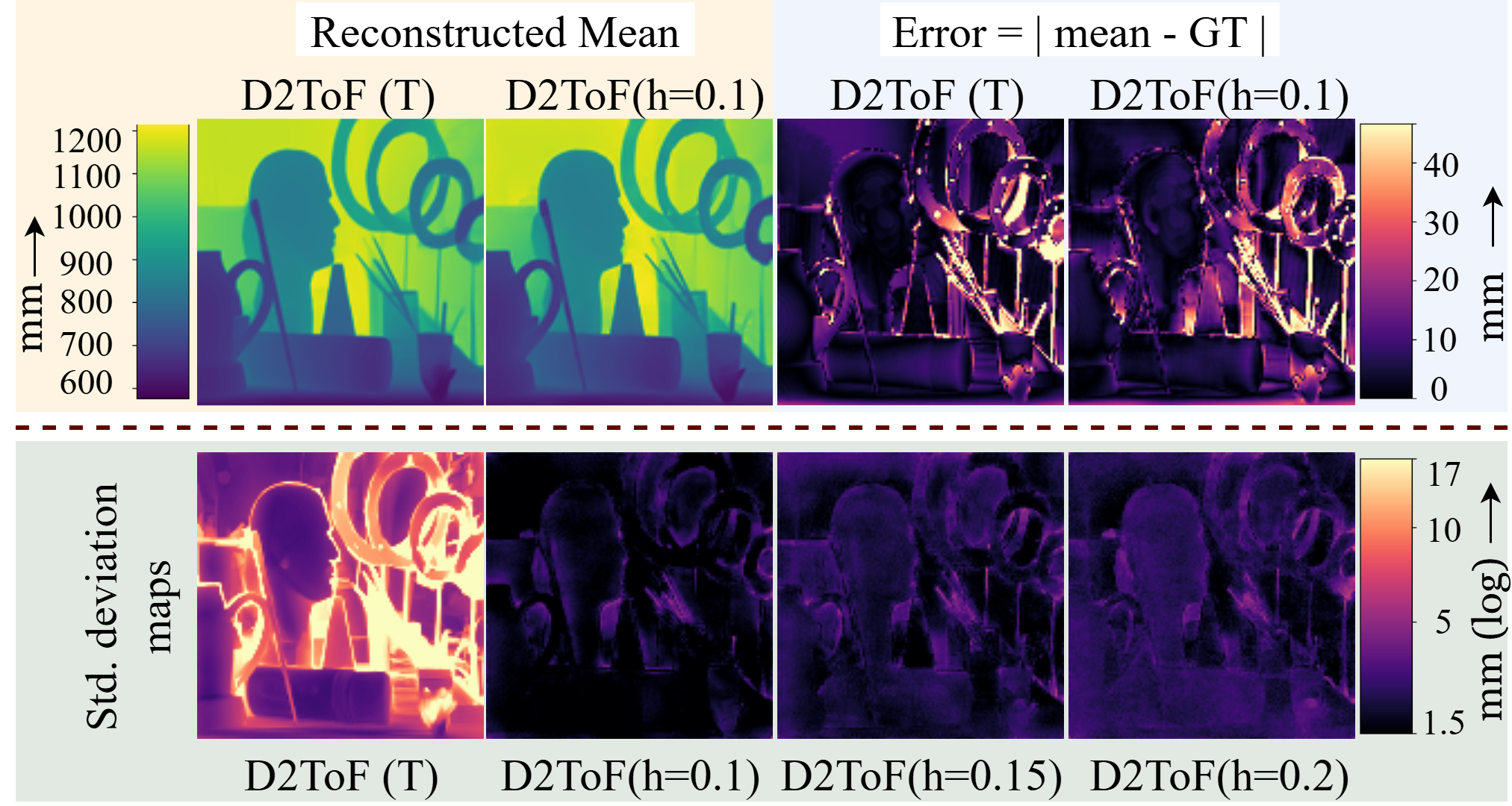}
    \caption{We show the uncertainty estimation (standard deviation) of our teacher and student model for the \textit{Art} scene from Middlebury in a \emph{clean} setting. D2ToF (T) denotes our teacher model.}
    \label{fig:ablation-uncertainty}
\end{figure}

Figure~\ref{fig:ablation-uncertainty} shows the per-pixel standard deviation maps of the teacher and the student in the clean setting, computed over $50$ samples drawn for the same input. The multi-step teacher exhibits a larger sample spread than the student with \(h=0.1\) {\br (in [0,1] space)}. For the student, increasing the PPB perturbation scale from \(h=0.1\) to \(h=0.2\) {\br (all separately trained)} correspondingly increases the variability of the reconstructed samples.

%% file: sections/05_conclusion.tex
\section{Conclusion} \label{sec:conclusion}

In this paper, we presented D2ToF, a one-step framework for extreme single-photon LiDAR depth super-resolution from $8\times8$ ToF histograms to $128\times128$ depth maps. Our approach distils a multi-step conditional diffusion teacher into a single-step generator using an amortized variational distillation framework. A differentiable measurement-constrained data-consistency term further encourages the reconstructed depth to reproduce the measured low-resolution ToF histogram under the forward model. To address the severe
ill-posedness of the reconstruction problem, we combine the ToF waveform with a high-resolution reflectivity image and an unaligned monocular relative-depth estimate, providing complementary metric and spatial guidance.
Our experiments demonstrate that the proposed approach outperforms the evaluated baselines across both clean and degraded measurement conditions, while the ablation studies demonstrate the benefits of the multimodal conditioning and histogram data-consistency term. Qualitative experiments on real $8\times8$ SPAD measurements further demonstrate transfer from simulated training data to real sensor captures under both well-lit and low-light conditions. The one-step model reconstructs a $128\times128$ depth map in approximately $83$~ms on an NVIDIA RTX~3090. Overall, D2ToF provides an efficient approach for high-resolution depth reconstruction from extremely low-resolution single-photon ToF measurements.